\documentclass[aps,prx,reprint]{revtex4-2}

\usepackage{etoolbox}
\usepackage{hyperref}
\usepackage{amsmath}
\usepackage{amssymb}
\usepackage{dsfont}
\usepackage[T1]{fontenc}
\usepackage{ae}
\usepackage{color}
\usepackage{graphicx} 
\usepackage{placeins}
\usepackage[mediumspace,mediumqspace,squaren]{SIunits} 

\newcommand{\ket}[1]{\ensuremath{\left| #1 \right\rangle}}
\newcommand{\bra}[1]{\ensuremath{\left\langle #1 \right|}}
\newcommand{\orho} {{\hat{\rho}}}
\newcommand{\dd} {\hbox{\textrm d}}
\newcommand{\oa} {{\hat{a}}}
\newcommand{\oad} {{\hat{a}^\dag}}
\newcommand{\os} {{\hat{\sigma}}}

\begin{document}

\title{An Intracavity Rydberg Superatom for Optical Quantum Engineering:\\
Coherent Control, Single-Shot Detection and Optical $\pi$ Phase Shift}

\author{Julien Vaneecloo} \affiliation{JEIP,  USR 3573 CNRS, Coll{\`e}ge de France, PSL University, 11, place Marcelin Berthelot, 75231 Paris Cedex 05, France}

\author{S{\'e}bastien Garcia} \affiliation{JEIP,  USR 3573 CNRS, Coll{\`e}ge de France, PSL University, 11, place Marcelin Berthelot, 75231 Paris Cedex 05, France}

\author{Alexei Ourjoumtsev} \email{Corresponding author: alexei.ourjoumtsev@college-de-france.fr} \affiliation{JEIP,  USR 3573 CNRS, Coll{\`e}ge de France, PSL University, 11, place Marcelin Berthelot, 75231 Paris Cedex 05, France}

\begin{abstract}
We demonstrate a new versatile building block for optical quantum technologies, based on an intracavity Rydberg-blockaded atomic ensemble acting as a single two-level superatom. We coherently control its state and optically detect it in a single shot with a $95\%$ efficiency. Crucially, we demonstrate a superatom-state-dependent $\pi$ phase rotation on the light reflected from the cavity. Together with the state manipulation and detection, it is a key ingredient for implementing deterministic photonic entangling gates and for generating highly non-classical light states.
\end{abstract}

\maketitle

\section{Introduction}

Optical photons are a central physical resource in quantum technologies. Necessitating neither vacuum nor cryogenics to preserve their quantum features, they are essential for connecting distant quantum nodes and play a key role in quantum-enhanced sensing. However, assembling a quantum system capable of performing complex communication, calculation, simulation or sensing tasks requires strong coherent interactions between its components, which photons do not easily provide.

For solving this long-standing issue, the most explored route consists in coupling photons to a highly anharmonic quantum system, most often epitomized by a single atom or a quantum dot, with an optical response strongly dependent on the number of photons it interacts with. This approach is very successful in the microwave domain, where the required anharmonicity is provided by circular Rydberg atoms \cite{Haroche2006} or Josephson junctions \cite{Devoret2013} coupled to superconducting resonators with very high quality factors. At optical wavelengths, to compensate for the relatively low transition dipoles, light fields need to be enhanced by confining the photons inside very low volume and high finesse optical cavities \cite{Thompson1992,Reithmaier2004,Yoshie2004}. A remarkable decades-long research effort in this direction \cite{Reiserer2015,Lodahl2015} recently led to first deterministic realizations of an optical two-photon quantum logic gate and of highly non-classical optical Schr\"odinger's cat states \cite{Hacker2016,Hacker2019}. Besides the difficulties in minimizing optical losses, reaching strong, steady and reproducible couplings in such structures remains very challenging: by lack of optical access, real atoms are difficult to position precisely and to keep steady, while artificial atoms embedded in microstructures are strongly influenced by the surrounding substrate.

An alternative approach consists in injecting the photons in an atomic gas and transiently converting them into highly polarizable Rydberg polaritons \cite{Firstenberg2016,Murray2017}. This conversion is controlled by a laser beam driving the upper branch of a two-photon transition, creating an electromagnetically-induced transparency (EIT) effect. Strong dipolar interactions between Rydberg atoms shift the energy of a pair out of the two-photon resonance. As a result, each polariton becomes surrounded by a blockade sphere where the EIT vanishes \cite{Pritchard2010}, which can be used to turn a coherent light beam into a stream of anti-bunched photons \cite{Dudin2012,Peyronel2012,Maxwell2013} and realize photon-controlled optical transistors \cite{Gorniaczyk2014,Tiarks2014}. Away from the single-photon resonance this blockade effect modulates not only the dissipation but also the optical dispersion \cite{Parigi2012}, allowing one to create bound photonic states \cite{Firstenberg2013,Liang2018,Stiesdal2018} and to perform two-photon quantum logic operations \cite{Tiarks2018}. Unfortunately, collisions between the Rydberg electron and the surrounding ground-state atoms \cite{Schlagmueller2016} limit the maximal optical density per blockade sphere, which defines the transmission or phase contrast created by a single Rydberg polariton. A conditional phase shift of $\pi$, required for most quantum logic and non-classical light engineering tasks, is then accompanied by significant losses \cite{Tiarks2016}.

Here, we experimentally demonstrate that this goal can be reached with a platform combining the two previous approaches and enhancing the Rydberg blockade in a small atomic cloud with a single-ended medium-finesse optical cavity. We show that the cloud acts as a single Rydberg superatom with a collectively enhanced coupling to light, which we can coherently manipulate and optically detect in a single shot with a $95\%$ efficiency via the transmission of the cavity. Most importantly, with respect to recent experiments on Rydberg superatoms \cite{Jia2018,Xu2021,Yang2021,Spong2021}, we successfully unlocked a qualitatively new regime where the phase of the light reflected from the cavity is shifted by $\pi$ by a single Rydberg excitation, allowing us to detect the latter with a $90\%$ efficiency via a homodyne measurement. This $\pi$ phase rotation, together with the coherent control and the single-shot state detection, is crucial for implementing two-photon quantum gates \cite{Das2016} and for generating non-classical optical resources for quantum sensing and communications.

\section{Strong superatom-photon coupling}

\begin{figure}[t]
\centering
\includegraphics[width=85mm]{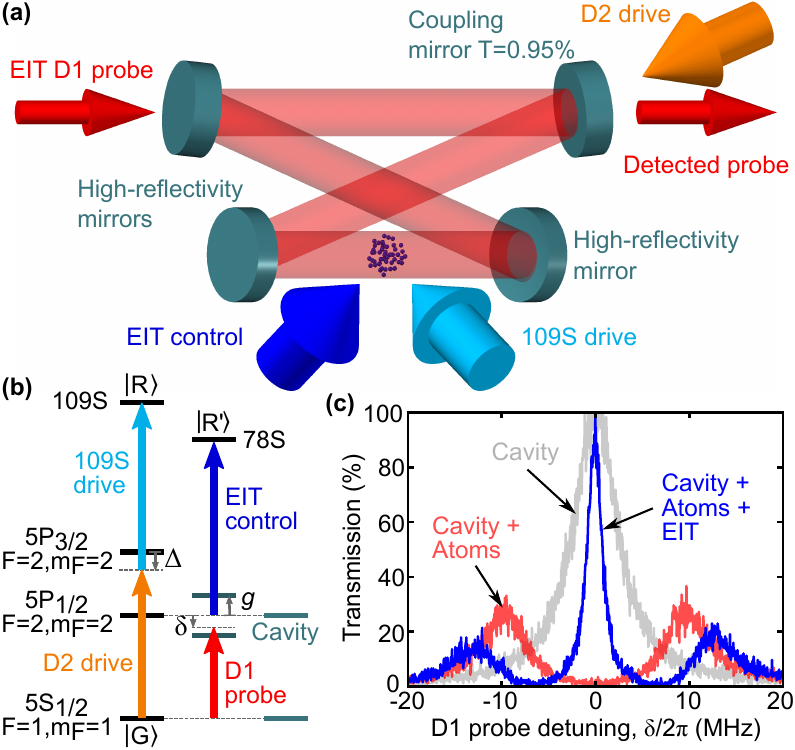}
\caption{(a) Experimental setup, configured for transmission measurement, and (b) level scheme: a small cloud of rubidium atoms behaves as a two-level superatom thanks to Rydberg blockade. D2 drive and 109S drive beams allow the excitation of the superatom to the \ket{R} state. The atomic ensemble is strongly coupled to a medium-finesse cavity, resonant with the D1 line. The control beam enables the transmission of the D1 probe through the cavity by Rydberg EIT with the 78S state. (c) Cavity EIT spectra: the transmission of the D1 probe, as a function of the probe detuning is measured with a single photon counting module (SPCM) and normalized to the maximum transmission of the empty cavity (light grey). With the strongly coupled atomic ensemble, we observe the vacuum Rabi splitting of the two transmission peaks, located at $\pm g/(2\pi)=\pm\unit{10}{\mega\hertz}$ (red). Adding the EIT control beam opens a transmission windows at resonance (blue). }
\label{fig:setup}
\end{figure}

The central feature of our experimental setup is a small cold $^{87}$Rb ensemble in an in-vacuum medium-finesse cavity, depicted in Fig.~\ref{fig:setup}(a). This ensemble, with a Gaussian rms radius of $\sigma_{a} = \unit{5}{\micro\meter}$, containing approximately $N = 800$ atoms at a temperature of \unit{3}{\micro\kelvin}, is prepared by sequentially cooling a background vapor in a 2-dimensional magneto-optical trap (2D MOT), loading a 3D MOT, transporting atoms with a 1D moving optical lattice towards the cavity mode, further cooling by 3D degenerate Raman sideband cooling, loading in a crossed dipole trap and pumping to the ground level $G: 5S_{1/2}, F\!\!=\!\!1, m_F\!\!=\!\!1$ (details in Appendix~\ref{app:CloudPrep}). A \unit{780}{\nano\meter} $D2$ and a \unit{480}{\nano\meter} $109S$ beams resonantly drive two-photon transitions to the Rydberg state $R: 109S_{1/2},J\!\!=\!\!1/2,m_J\!\!=\!\!1/2,I\!\!=\!\!3/2,m_I\!\!=\!\!3/2$ with a detuning $\Delta/(2\pi) = -\unit{545}{\mega\hertz}$ below the intermediate state $5P_{3/2}, F\!\!=\!\!2, m_F\!\!=\!\!2$. Interactions between Rydberg atoms, characterized at long range by the van-der-Waals coefficient $C_{RR} = \unit{154}{\tera\hertz.\micro\meter^6}$, shift the resonance of a Rydberg pair by $C_{RR}/r^6 > \unit{2.4}{\mega\hertz}$ for atoms separated by $r < 4 \sigma_{a}$, a condition satisfied by more than $95\%$ of atom pairs. This blockade of multiple Rydberg excitations makes the cloud behave as a two-level superatom with a ground state \ket{G} and a state \ket{R} corresponding to a single delocalized Rydberg $109S$ excitation.

The atoms are positioned at the \unit{21}{\micro\meter} waist of a circularly-polarized running-wave nearly-Gaussian mode, defined by a single-ended 4-mirror non-planar cavity with a finesse $\mathcal{F}=590$. This mode, driven by coherent laser pulses, is resonant with the $D1$-line transition from the ground state to the $5P_{1/2}, F\!\!=\!\!2, m_F\!\!=\!\!2$ level, see Fig.~\ref{fig:setup}(b-c). The coupling mirror has a $0.95\%$ transmission allowing $90\%$ of the photons to leave the cavity through this port. The atomic ensemble is collectively coupled to the cavity mode with a constant $g = 2\pi \times \unit{10}{\mega\hertz}$ exceeding the decay rates of both the cavity field ($\kappa = 2\pi \times \unit{2.9}{\mega\hertz}$) and the atomic dipole ($\gamma=2\pi\times\unit{3}{\mega\hertz}$), leading to a vacuum Rabi splitting visible on the transmission spectrum (Fig.~\ref{fig:setup}(b: red curve)). By adding a control laser coupling the $5P_{1/2}$ state to the $R': 78S_{1/2},J\!\!=\!\!1/2, m_J\!\!=\!\!1/2, I\!\!=\!\!3/2, m_I\!\!=\!\!3/2$ Rydberg state, we open an EIT window on resonance (Fig.~\ref{fig:setup}(c: blue curve)) and recover $90\%$ of the bare cavity transmission for the $D1$ probe beam. This EIT window corresponds to a dark polariton state mixing a photon in the cavity with a Rydberg excitation \ket{R'} in the cloud. Atoms in states $R$ and $R'$ interact strongly via their dipoles, in particular because this pair of states is close to a F\"orster resonance with the $110P_{3/2}+77P_{3/2}$ pair. The asymptotic van-der-Waals scaling coefficient $C_{RR'}=\unit{18}{\tera\hertz.\micro\meter^6}$ overestimates the interactions in our parameter regime, but a complete diagonalization \cite{Sibalic2017} gives an interaction-induced frequency shift of \unit{3.4}{\mega\hertz} at the most likely distance $r=2 \sigma_{a}$. Therefore, a superatom in \ket{R} efficiently shifts \ket{R'} out of resonance, destroying the EIT and strongly changing the response of the system to the $D1$ probe which drives the photonic component of the \ket{R'} polariton.

\FloatBarrier

\section{Control and single-shot detection}

\begin{figure}[t]
\centering
\includegraphics[width=85mm]{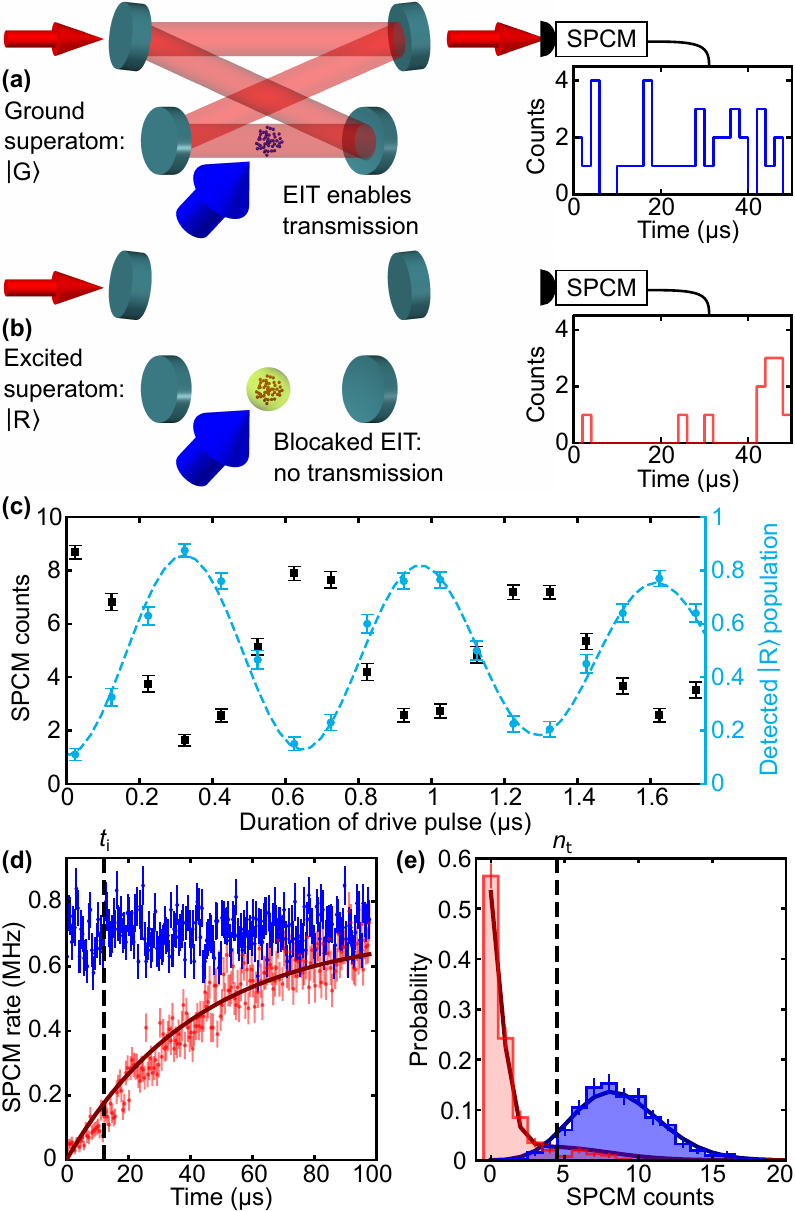}
\caption{ Detection in transmission: the photon flux, detected with an SPCM, switches from (a) high without 109S Rydberg excitation to (b) quasi zero with a single 109S excitation blockading the EIT (blockade sphere depicted in yellow). The insets show examples of single-shot measures with \unit{2}{\micro\second} binning. (c) 109S Rabi oscillations are visible on the averaged SPCM counts in the detection windows (black) and the 109S population deduced from single-shot detections (cyan). Each data point represents $200$ detections. The dashed line is a best-fit Gaussian decoherence model. (d) SPCM count rate since the beginning of probe D1 pulse without (blue) and with (red) a 109S Rydberg drive $\pi$ pulse. The dark-red line is a best-fit exponential decay with a lifetime $\tau_r = \unit{42}{\micro\second}$. The dashed vertical line marks out the optimized detection window $t_i$. (e) Histograms of the number of photons, without (blue) and with (red) $\pi$ pulse. The dark curves represent modeled histograms. The dashed vertical line shows the optimized detection threshold $n_t$. (d),(e): data are averaged over $400$ repetitions. All errorbars represent the standard error.
}
\label{fig:detSPCM}
\end{figure}

Fig.~\ref{fig:detSPCM} illustrates the single-shot detection of the superatom's state via the transmission of the $D1$ probe through the cavity. When the superatom is driven from \ket{G} to \ket{R}, the EIT vanishes and we observe a clear drop in the transmitted photon flux, detected with a fiber-coupled single photon counting module (SPCM) with a total detection efficiency of $47\%$ (insets of Fig.\ref{fig:detSPCM}.a-b). This flux can then be used to monitor Rabi oscillations of the superatom's state, driven with time-centered $D2$ and $109S$ beam pulses (Fig.~\ref{fig:detSPCM}c). The former, shorter than the latter by \unit{0.2}{\micro\second}, set the effective excitation duration $t_d$. \unit{1}{\micro\second} after the excitation finishes, we switch on the $D1$ cavity probe and integrate the transmitted photon detections over a time $t_i$ optimized as discussed below. From these single-shot results, we infer the statistic \ket{R} population plotted in Fig.~\ref{fig:detSPCM}(c), where the probe power was set to obtain a \unit{350}{\kilo\hertz} count rate for a superatom in \ket{G} and $t_i=\unit{24}{\micro\second}$. The dashed curve is a best-fit by a sine function with a Gaussian decaying envelope, yielding the 2-photon Rabi frequency $\Omega/(2\pi) = \unit{1.5}{\mega\hertz}$ and the $e^{-1}$ Gaussian decay time $\tau_d = \unit{2.8}{\micro\second}$. The value of the Rabi frequency matches the expected behavior $\Omega = \sqrt{N} \Omega_{D2} \Omega_{109S} / (2 |\Delta|)$, where the respective single-photon Rabi frequencies of the D2 and 109S drives $\Omega_{D2}/(2\pi) = \unit{6}{\mega\hertz}$ and $\Omega_{109S}/(2\pi) = \unit{10}{\mega\hertz}$ were determined independently and the $\sqrt{N}$ scaling \cite{Dudin2012b} was verified by partially depumping the atoms to the dark manifold $5S_{1/2},F=2$ and determining $N$ from the vacuum Rabi splitting. The decay time $\tau_d$ is consistent with a motional dephasing of the polariton at \unit{3}{\micro\kelvin} combined with $4\%$-rms fluctuations of $\Omega$ dominated by fluctuations on $N$, with calculated Gaussian decay times of \unit{3.8}{\micro\second} and \unit{3.7}{\micro\second}, respectively.

Once the Rabi $\pi$-pulse is calibrated, we can optimize the state detection. We first observe that the \ket{R} state has a limited lifetime, visible as quantum jumps on individual SPCM records (see Fig.~\ref{fig:detSPCM}(b:inset) at \unit{42}{\micro\second}) and as an exponential evolution on the mean rate (Fig.~\ref{fig:detSPCM}(d:red)). This effect limits the detection duration $t_i$ because a quantum jump before $t_i$ may induce a false-negative error. Increasing the input photon flux could make the detection faster, but we observed that the excited-state lifetime decreases proportionally. This decrease may result from light-assisted collisions, difficult to model since each $109S$ atom has $\sim 13$ other atoms within the  \unit{2.8}{\micro\meter} LeRoy radius below which the $109S$ and $78S$ wavefunctions overlap significantly. This inverse proportionality relation effectively limits the number of photon counts providing a sufficiently low false-positive error probability $\epsilon_r$. The photon flux will ultimately be limited by the self-blockade of \ket{R'} polaritons ($C_{R'R'}=\unit{3}{\tera\hertz.\micro\meter^6}$) and saturate towards $\sim 1/(2\tau_p) = \unit{6}{\mega\hertz}$, with $\tau_p= \unit{85}{\nano\second}$ the polariton lifetime, extracted from the EIT linewidth in Fig.~\ref{fig:setup}(c).

The self-blockade makes the transmitted flux sub-Poissonian, with a measured zero-delay autocorrelation function $g^{(2)}(0) = 0.15$, but we recover $g^{(2)}(\tau) = 1$ for $\tau \gg \tau_p$. As we integrate the photon counts for at least a few microseconds, their statistic is quasi Poissonian. Therefore, in Fig.~\ref{fig:detSPCM}(e), the histogram of single-shot SPCM counts $n$ in a $t_i =\unit{12}{\micro\second}$ window with a ground-state superatom (blue) obeys $\mathcal{P}(n, t_i \phi_G)=e^{-t_i \phi_G}(t_i \phi_G)^n/n!$ (dark blue) with a mean count number of $t_i \phi_G = 8.7$ where $\phi_G$ is the average flux. When we apply a $\pi$ pulse to excite the superatom to \ket{R}, the corresponding histogram (red) shows that most of the single-shot detections result in zero SPCM counts, as expected from EIT blockade. The residual transmission results from three effects. First, finite-strength atom-light couplings and Rydberg interactions leave a residual transmitted flux $\phi_R$. Second, even though the \ket{R} lifetime $\tau_R = \unit{42}{\micro\second}$ largely exceeds $t_i$, quantum jumps to \ket{G} occur within the detection window with a $25\%$ probability. Third, the preparation efficiency $\eta_R$ of \ket{R}, limited to $e^{-t_{\pi}^2/\tau_d^2} = 99\%$ by the duration $t_{\pi}$ of the $\pi$ pulse, contributes to the measured histogram. Thus, on the fitted model (dark red, see Appendix~\ref{app:DetModels}) that accounts for these effects, we leave $\phi_R$ and $\eta_R$ as free parameters, yielding best-fit values $\phi_R/\phi_G = 4.5\pm 0.4 \%$ and $\eta_R = 100^{+0}_{-5}\%$. 

\begin{figure}[t]
	\centering
	\includegraphics[width=85mm]{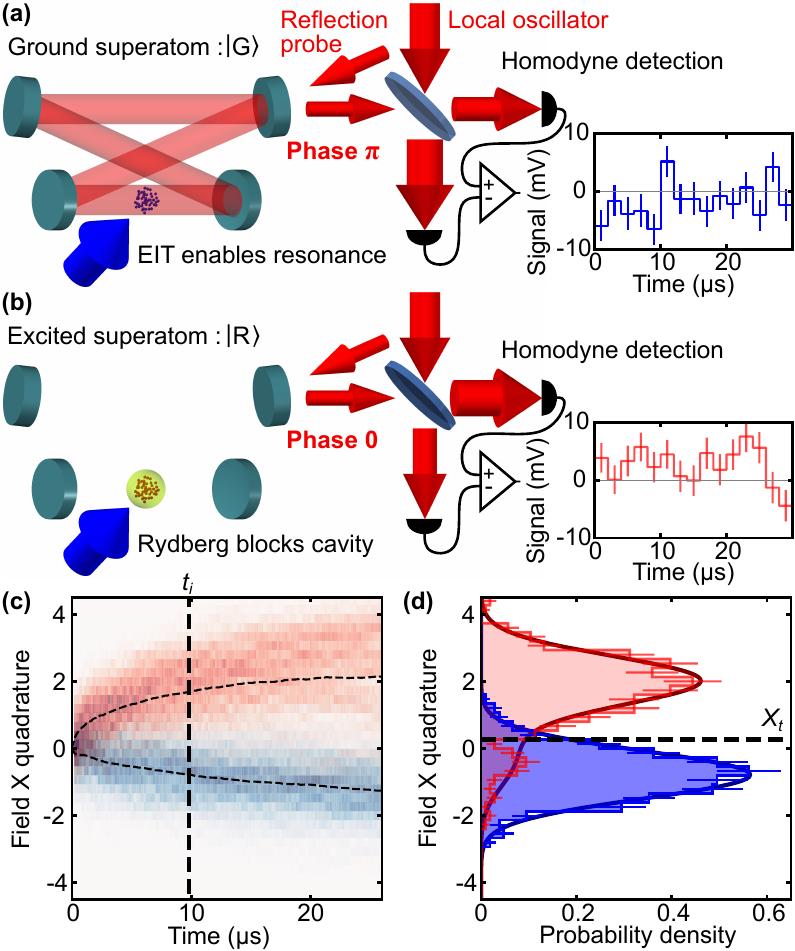}
	\caption{ $\pi$ phase shift in reflection: the reflected probe is combined with a local oscillator beam on a beamsplitter, whose output intensity difference is measured by an HD setup.  With the addition of a 109S Rydberg excitation blockading the EIT, the reflected probe phase changes from $\pi$ (a) to $0$ (b), resulting in a sign flip of the measured HD signal. The insets show examples of single-shot measures with a \unit{2}{\micro\second} binning. (c) In-phase quadrature X since the beginning of probe D1 pulse without (blue) and with (red) a 109S Rydberg drive $\pi$ pulse.  The dashed lines indicates the quadrature means and the color represent the measured probability density. The dashed vertical line gives the upper limit of the optimized detection window. (d) Histograms of the field X quadrature values, without (blue) and with (red) $\pi$ pulse. The dark curves represent modeled histograms. The dashed vertical line shows the optimized detection threshold. (c),(d): data are averaged over $400$ repetitions. All errorbars represent the standard error.}
	\label{fig:detHD}
\end{figure} 

Setting a detection threshold $n_t = 5$ (dashed line) in between the two distributions, with and without a $\pi$ pulse, allows us to infer the most probable state of the superatom from a single-shot SPCM count $n$: \ket{G} if $n \geq n_t$ or \ket{R} if $n < n_t$. The integrals of the histograms, taken over $n \geq n_t$ with a $\pi$ pulse and over $n < n_t$ without, give respectively the false-negative error $\epsilon_R = 4.8 \pm 1.1\%$ and the false-positive error $\epsilon_G = 5.3 \pm 1.1\%$. A conservative definition of the detection fidelity gives $\mathcal{F} = 1 - max(\epsilon_G,\epsilon_R)= 94.7\pm 1.1\%$, optimized by varying $t_i$ and $n_t$. The resulting $t_i=\unit{12}{\micro\second}$ is a trade-off: short times avoid errors induced by quantum jumps, while long times reduce errors through a lower uncertainty on the photon flux.

\section{Optical $\pi$ phase shift}

A cavity is optically far-overcoupled when the transmission of the input/output coupler greatly exceeds all other losses, and far-undercoupled in the opposite limit case. A resonant impinging beam is efficiently reflected in both cases, but with a $\pi$ phase difference. In our case, for a superatom in \ket{G} the EIT is good enough for the system to be overcoupled, while in \ket{R} the blockade is sufficiently strong to switch the optical impedance to a far-undercoupled regime. We observe the expected $\pi$ phase flip on a $D1$ probe reflected from the cavity using a balanced homodyne detection (HD) with an overall 75\% efficiency, see Fig.~\ref{fig:detHD}. The phase of the probe relative to the local oscillator (LO) is stabilized by a lock-and-hold technique and referenced to a far-off-resonant cavity. For a superatom in \ket{G}, the system is resonantly excited and we extract a phase $\theta_{G}=\unit{\pi(0.91\pm 0.02)}{\radian}$, while for \ket{R} the system becomes off-resonant and we obtain $\theta_{R}=\unit{\pi(-0.045\pm 0.013)}{\radian}$, yielding a phase shift $\theta_{G}-\theta_{R}= \unit{\pi(0.96\pm 0.03)}{\radian}$.

As a first application, we use this phase flip to detect the superatom's state, via single-shot homodyne measurements of the in-phase quadrature $\hat{X}$ shown in Fig.~\ref{fig:detHD}(c), with the vacuum noise variance normalized to $1/2$. The input photon rate is at \unit{580}{\kilo\hertz}, on the edge of self-saturation effects, for which we observe a \unit{38}{\micro\second} Rydberg lifetime. We reach a $89.9\pm 1.5\%$ fidelity from the two measured density distributions in Fig.~\ref{fig:detHD}(d), for a \unit{10}{\micro\second} integration time and a threshold set to X$_t$=0.27. The fidelity is limited by the false-negative errors while the false-positive probability value is $6.9\pm 1.3 \%$.
We model these distributions like previously (see Sup. Mat. B), taking into account the Rydberg state decay, leaving as free parameters the preparation efficiency $\eta_R=99^{+1}_{-2}\%$ and the reflectivity level for the superatom in the Rydberg state (red) $\mathcal{R}_{R} = 51 \pm 2 \%$.  The distribution for \ket{G} is fitted by a Gaussian centered on the measured mean value (blue).  
This approach is more sensitive to probe self-saturation effects which directly decreases the HD signal. Here, the self-blockade yields a $\sim 7\%$ effective reflectivity for \ket{G} compared to $\mathcal{R}_{G}=43\pm 3\%$ in the low-intensity linear regime (see Appendix~\ref{app:EITSpectra}).

\FloatBarrier

\section{Conclusion and perspectives}

In conclusion, the demonstrated coherent state control, the two single-shot state detection methods and the conditional optical $\pi$ phase shift confirm that combining a Rydberg superatom with a medium-finesse cavity is a viable route towards deterministic optical quantum engineering. The state detection efficiencies match those recently obtained in Rydberg-based experiments operating in free space or in low-finesse resonators \cite{Xu2021,Yang2021}, but the $\pi$ phase flip is a unique feature, relying on an excellent EIT combined with a strong Rydberg blockade, that would vanish for a low- or high-finesse cavity and would require post-selection for free-space beams \cite{Tiarks2016,Beck2016}.

In comparison with the well-explored single-atom cavity QED route, our approach is still in its infancy. Nevertheless, it already offers a similar photon extraction efficiency \cite{Hacker2019} while having a considerable margin for improvement, all the more as in our setup we chose versatility over performance. For instance, cavity losses could be reduced by at least an order of magnitude by using superpolished mirrors with less broadband coatings and by cleaning the deposits that formed during the initial testing of the setup. Taking all sources of decoherence into account we estimate that this improvement alone should increase the expected Choi-Jamio{\l}kowski fidelity of a two-photon control-Z gate feasible with this setup from $\approx 80\%$ to $\approx 95\%$ \cite{Das2016}. The strong collective couplings of Rydberg-blockaded atomic ensembles allow the use of large cavities with non-trivial cavity geometries such as ours, which recently led to the first observation of optical Laughlin states \cite{Clark2020}. The Rydberg blockade mechanism could also be used to make neighboring superatoms interact with each other \cite{ParedesBarato2014,Khazali2019}, opening countless possibilities for multimode quantum optics and quantum simulations.

\section*{Acknowledgments}

This work was funded by the IDEX grant ANR-10-IDEX-001-02-PSL PISE and the ERC Starting Grant 677470 SEAQUEL. The authors thank P. Travers for technical support, S. \'Cuk and M. Enault-Dautheribes for their assistance at the early stage of the project, and V. Magro for his help during characterization measurements. A.O. is a CIFAR Azrieli Global Scholar.

\appendix

\section{Atomic cloud preparation and driving}
\label{app:CloudPrep}

The source of atoms is a sample of metallic rubidium heated at $37$°C, providing a vapor pressure on the order $10^{-7}\,$mBar in a side vacuum chamber. In this side chamber, we create a two-dimensional magneto-optical trap (2D MOT) to cool $^{87}$Rb with \unit{50}{\milli\watt} of cooling laser light (red detuned by \unit{-10}{\mega\hertz} from the transition $5S_{1/2}, F=2 \rightarrow 5P_{3/2},  F=3$) and \unit{0.4}{\milli\watt} of repumping laser light (resonant with the transition $5S_{1/2}, F=1 \rightarrow 5P_{3/2}, F=2$). With a part of the cooling light, we push the cold atoms through a differential pumping tube into the main vacuum chamber containing the rest of the setup at a pressure of \unit{2\,10^{-9}}{\milli\bbar}. The atomic beam loads a 3D MOT at a rate of $7\, 10^5$ atoms per ms. The 3D MOT is formed by $3$ pairs of orthogonal beams, with each a waist of \unit{7.5}{\milli\meter}, \unit{12}{\milli\watt} of cooling light (detuned by \unit{-18}{\mega\hertz}) and \unit{0.2}{\milli\watt} of repumping light. The \unit{8}{\giga\per\centi\meter} magnetic field gradient is generated by a pair in-vacuum coils in anti-Helmholtz configuration. After \unit{50}{\milli\second} of loading, we compress the 3D MOT for \unit{1}{\milli\second} by ramping up the magnetic field to \unit{21}{\giga\per\centi\meter}. We follow with a molasses phase for \unit{2}{\milli\second}, during which we ramp down the magnetic field and the beams intensities while increasing the detuning of the cooling light to \unit{-70}{\mega\hertz}.  

The center of the 3D MOT is located \unit{33}{\milli\meter} below the waist of the cavity mode. To transport atoms over this distance, we use a 1D-optical-lattice dipole trap as a conveyor belt for atoms~\cite{Schrader2001}. The lattice is formed with two vertical counter-propagating laser beams at an identical wavelength of approximately \unit{782.9}{\nano\meter}, each with a power of \unit{0.25}{\watt} and a waist of \unit{60}{\micro\meter} located midway between the 3D MOT and the cavity mode. We start by loading the lattice with atoms at the 3D MOT position during the molasses step. By linearly sweeping the frequency of one of the beams up then down, we displace the interference fringes and thus the atoms trapped in maxima of light intensity. In \unit{10}{\milli\second}, we transport to the cavity center an ensemble of $\sim 1$ million atoms with a temperature of about \unit{40}{\micro\kelvin}. 

The thermal atomic motion has several detrimental effects. First, the Doppler broadening increases the transition linewidth to Rydberg levels which reduces the EIT transparency. Second, the higher the temperature the deeper a dipole trap has to be, thus the differential lightshift over the trapped cloud also broadens the transition linewidth to Rydberg levels. And third, the coherence of the Rydberg excited state of the superatom relies on the relative phases $\vec{k}_{GR}\cdot \vec{r}_i$ impinged by the exciting lasers, with summed wavevector $\vec{k}_{GR}$, onto the atom $i$ at position $\vec{r}_i$. The thermal motion thus induces a Gaussian decay of the coherence $e^{-t^2/\tau_{d,T}^2}$ at a characteristic time $\tau_{d,T} = \sqrt{m/k_{\mathrm{B}} T} / \left\|\vec{k}_{GR}\right\|$, where $m$ is the mass of the atom, $k_{\mathrm{B}}$ is the Boltzmann constant and $T$ is the temperature of the cloud. 

In order to limits these effects, we implement degenerate Raman sideband cooling~\cite{Hamann1998,Kerman2000} once the atoms are at the cavity position. For this, we use a 3D optical lattice formed on vertical axis with the conveyor belt lattice and in the horizontal plane by two independent and orthogonal interference patterns. The latter two are formed by two laser beam pairs whose waists are \unit{0.17}{\milli\meter} and whose frequencies differ by \unit{440}{\mega\hertz} and are blue detuned by \unit{16}{\giga\hertz} from the transition $5S_{1/2} F=2 \rightarrow 5P_{3/2} F=3$. For one pair, the angle between the beam's wavevectors is \unit{90}{\degree}, while for the other pair the angle is only \unit{43}{\degree}. Nevertheless, by adjusting the beams powers to \unit{\approx 3}{\milli\watt} and \unit{\approx 10}{\milli\watt} respectively, we obtain isotropic trapping sites whose oscillation frequency is \unit{0.2}{\mega\hertz}. To do so, we measured the oscillation frequencies of each of the three orthogonal lattices by observing parametric heating losses when we modulate the beams intensities. The isotropic oscillation frequency allows us to perform degenerate Raman sideband cooling in 3D. A magnetic field of \unit{0.2}{\giga} brings to degeneracy the states $5S_{1/2}, F=2, m_F=2 , n$ and $5S_{1/2} F=2, m_F=1 , n-1$ with $n$ the motional quantum number. Small angles of \unit{\sim 10}{\degree} between the linear polarisations in each pair of beams induce Raman transitions from $5S_{1/2}, F=2, m_F=2 , n$ to $5S_{1/2}, F=2, m_F=1 , n-1$. A $\sigma^{+}$-polarized  Zeeman pumping beam with a \unit{1}{\milli\meter} waist and \unit{2}{\micro\watt} of power drives the transition $5S_{1/2}, F=2 \rightarrow 5P_{3/2}, F=2$ and, in presence of a hyperfine repumper from $F=1$, brings the atoms back to the dark $m_F=2$ state. This transfer back to $m_F=2$ conserves the vibrational quantum number because we operate in the Lamb-Dicke regime where the recoil frequency of \unit{\approx 10}{\kilo\hertz} is much smaller than the trap frequency of the lattice. The combination of Raman transitions and Zeeman pumping thus actively cools the atoms along all three dimensions. By applying this cooling for \unit{5}{\milli\second} to the atomic cloud inside the cavity, we reach a temperature of \unit{1}{\micro\kelvin}. Several elements are crucial for minimizing this temperature. Firstly, we pay special attention to the efficiency of the Zeeman pumping by using a clean circular polarization and carefully aligning the magnetic field along the beam axis using microwave spectroscopy between the $F=1$ and $F=2$ hyperfine manifolds of the $5S_{1/2}$ ground state. Secondly, our horizontal lattice beams are relatively close to the D2 transition in order to get large dipole trapping frequencies with moderate beam intensities. Here blue-detuned beams proved to be much more efficient for cooling than red-detuned ones, mainly because the trapping sites are at light intensity minima, thereby strongly reducing spontaneous emission. In addition, when atoms are trapped in near darkness, AC Stark shifts become negligible and do not affect the frequency of the Zeeman pump.

To achieve efficient Rydberg blockade, we reduce the cloud's size by loading it in a crossed dipole trap formed by two orthogonal laser beams with \unit{\approx 20}{\milli\watt} of power at a \unit{1064}{\nano\meter} wavelength with a \unit{220}{\mega\hertz} frequency difference to avoid interference. Two in-vacuum aspheric lenses with \unit{50}{\milli\meter} focal lengths and numerical apertures of $0.2$ focus the two beams to \unit{\sim 15}{\micro\meter} waists at the cavity mode center. The trap loading is performed by alternating Raman cooling steps with compression steps, where we let cold atoms fall towards the trap center. In between these steps, we ramp up or down adiabatically the 3D lattice used for Raman cooling within \unit{0.1}{\milli\second}. After a final step of Raman cooling, we wait for \unit{50}{\milli\second} to let the cloud thermalize, which removes residual atoms with mechanical energies of the order of the trap depth. We end up with a cloud having an rms radius of \unit{5}{\micro\meter}, containing about $800$ atoms at a peak density of \unit{4\,10^{11}}{\centi\meter\rpcubed} and a temperature of \unit{3}{\micro\kelvin}. 

Before performing the experiments presented in the main text, the final step is to optically pump the cloud into the state \ket{G} where the atoms are in the state $5S_{1/2}, F\!\!=\!\!1, m_F\!\!=\!\!1$. At the beginning of this step, we apply for \unit{2}{\micro\second} a beam resonant with the transition $5S_{1/2}, F=2 \rightarrow 5P_{3/2}, F=2$ that pumps the atoms to $F=1$. A Zeeman pumping beam pumps them to $m_F=1$ during \unit{100}{\micro\second} by driving a $\sigma^{+}$ transition $5S_{1/2}, F=1 \rightarrow 5P_{3/2}, F=1$. The Zeeman pumping beam is circularly polarized and sent through the science cavity ($\sim$\unit{200}{\mega\hertz} out of a resonance) in the opposite direction to the probing beam used later on. A \unit{3}{\giga} quantization magnetic field is applied with its direction quasi aligned with the cavity mode wavevector at the position of the atoms; the residual \unit{12}{\degree} angle stems from our beam geometry and ensures $\pi$-polarized transitions to Rydberg states. We performed microwave spectroscopy within the $5S_{1/2}$ manifold and found that the targeted $m_F=1$ state is populated at $95\pm2\%$, while the population in $m_F=0$ is $5\pm2\%$ and the population in $m_F=-1$ is negligible. With respect to $m_F = 1$, the collective coupling strength of the population in $m_F = 0$ is reduced by an order of magnitude by lack of collective enhancement and due to a smaller transition dipole moment. Thus, we can consider the ensemble as efficiently described by a superatom in state \ket{G}.     

To excite the superatom to \ket{R}, we use a $D2$ drive circularly-polarized beam that is spatially mode-matched and spectrally \unit{200}{\mega\hertz} out of resonance with the fundamental mode of the science cavity, with a power of \unit{2}{\micro\watt}. The $109S$ driving beam, with a waist of about \unit{60}{\micro\meter} and a power of \unit{4}{\watt}, is produced in an in-vacuum build-up cavity in confocal configuration. The build-up cavity fundamental transverse mode is located \unit{0.5}{\milli\meter} below the atomic cloud and it allows us to lock the cavity on resonance with a beam detuned by one free spectral range (\unit{0.5}{\giga\hertz}). The $109S$ driving beam axis is orthogonal to the \unit{3}{\giga} quantization magnetic field and its polarization is aligned with the field, thus driving $\pi$ transitions only towards Rydberg states. The $D2$ and $109S$ driving pulses are time-centered and switched on and off using acousto-optic modulators (AOMs). Because the high-power $109S$ beam is less tightly focused in the AOM than the $D2$ probe, its switching time is slower. Therefore, we make the $109S$ pulse \unit{0.2}{\micro\second} longer than the $D2$ pulse which sets the variable effective duration of the 2-photon Rabi drive.

The $78S$ EIT control beam is produced with another in-vacuum build-up cavity, with a similar locking scheme, a similar waist, and a power of \unit{5}{\watt}. 

\section{Models for the detections}
\label{app:DetModels}

Here, we present expressions of the fitting functions of the detections histograms.

We start with the model used to fit the data in Fig.~\ref{fig:detSPCM}(e) of the detection in transmission measured by the SPCM. When the superatom is in the ground state \ket{G}, the distribution of photon number $n$ is a Poisson law, $\mathcal{P}_G(n) = \mathcal{P}(n, t_i \phi_G)=e^{-t_i \phi_G}(t_i \phi_G)^n/n!$, parametrized by the mean photon number detected $t_i \phi_G$ with $\phi_G$ the photon rate and $t_i$ the integration time. For the superatom in the Rydberg state \ket{R}, the distribution can be split in two parts to take into accounts quantum jumps. First, the probability distribution without a jump is also a Poisson law $\mathcal{P}(n, t_i \phi_R)$ with a residual photon number $t_i \phi_R$, where $\phi_R$ is the photon flux with blockade. The probability of this event is $e^{-t_i/\tau_R}$, where $\tau_R$ is the Rydberg lifetime. For a quantum jump at $t<t_i$, the distribution is the sum of two Poisson laws, one with a mean photon number $t \phi_R$ and the second after the jump with a mean value $(t_i-t) \phi_G $. By the stability of the Poisson law under addition, this probability distribution is also a Poisson law with a mean photon number $n_J(t)= t \phi_R + (t_i-t) \phi_G$. As a result, the full distribution is: 
\begin{equation}
	\mathcal{P}_R(n) = \mathcal{P}(n,  t_i \phi_R)e^{-t_i/\tau_R} + \int_{0}^{t_i} \mathcal{P}(n, n_J(t)) \frac{e^{-t/\tau_R}}{\tau_R} dt
\end{equation}
Finally, the probability distribution with a preparation efficiency $\eta_R$ is given by $\eta_R \mathcal{P}_R(n)+(1-\eta_R)\mathcal{P}_G(n)$.

We now move to the fitting functions for the data in Fig.~\ref{fig:detHD}(d) where the superatom's state is measured by monitoring the field reflected off of the cavity using a homodyne detector. When the superatom lies in the ground state \ket{G}, the X quadrature is a Gaussian, $\mathcal{G}_G (X) = \mathcal{G} (X, \bar{X}_G)$, with a rms width $\sigma=1/\sqrt{2}$ and centered on $\bar{X}_G=-\sqrt{2 t_i \phi \mathcal{R}_G}$ , where $\phi$ is the photon flux at cavity input and $\mathcal{R}_G$ is the reflectivity when the superatom is in \ket{G}. In the absence of a quantum jump, the distribution associated to the superatom prepared in the Rydberg state is also a Gaussian centered on $\bar{X}_R=\sqrt{2 t_i \phi \mathcal{R}_R}$, where $\mathcal{R}_R$ is the associated reflectivity. For a quantum jump at $t<t_i$, the mean value of the quadrature is then $\bar{X}_J (t) = -(t_i-t) \sqrt{2 \phi \mathcal{R}_G / t_i } + t \sqrt{2\phi \mathcal{R}_{R} /t_i }$. Because it is the sum of two Gaussian random variables, the distribution remains Gaussian with a mean $\bar{X}_J (t)$ and a width of $1/\sqrt{2}$. The complete distribution for the superatom in \ket{R} is then:
\begin{equation}
		\mathcal{G}_R(X) = \mathcal{G}(X, \bar{X}_R) e^{-t_i/\tau_R} + \int_{0}^{t_i} \mathcal{G}(X, \bar{X}_J (t)) \frac{e^{-t/\tau_R}}{\tau_R} dt
\end{equation}
Finally, the probability distribution with a preparation efficiency $\eta_R$ is given by $\eta_R \mathcal{G}_R(n)+(1-\eta_R)\mathcal{G}_G(n)$.

\section{EIT transmission and reflection spectra}
\label{app:EITSpectra}

\begin{figure}[t]
	\centering
	\includegraphics[width=70mm]{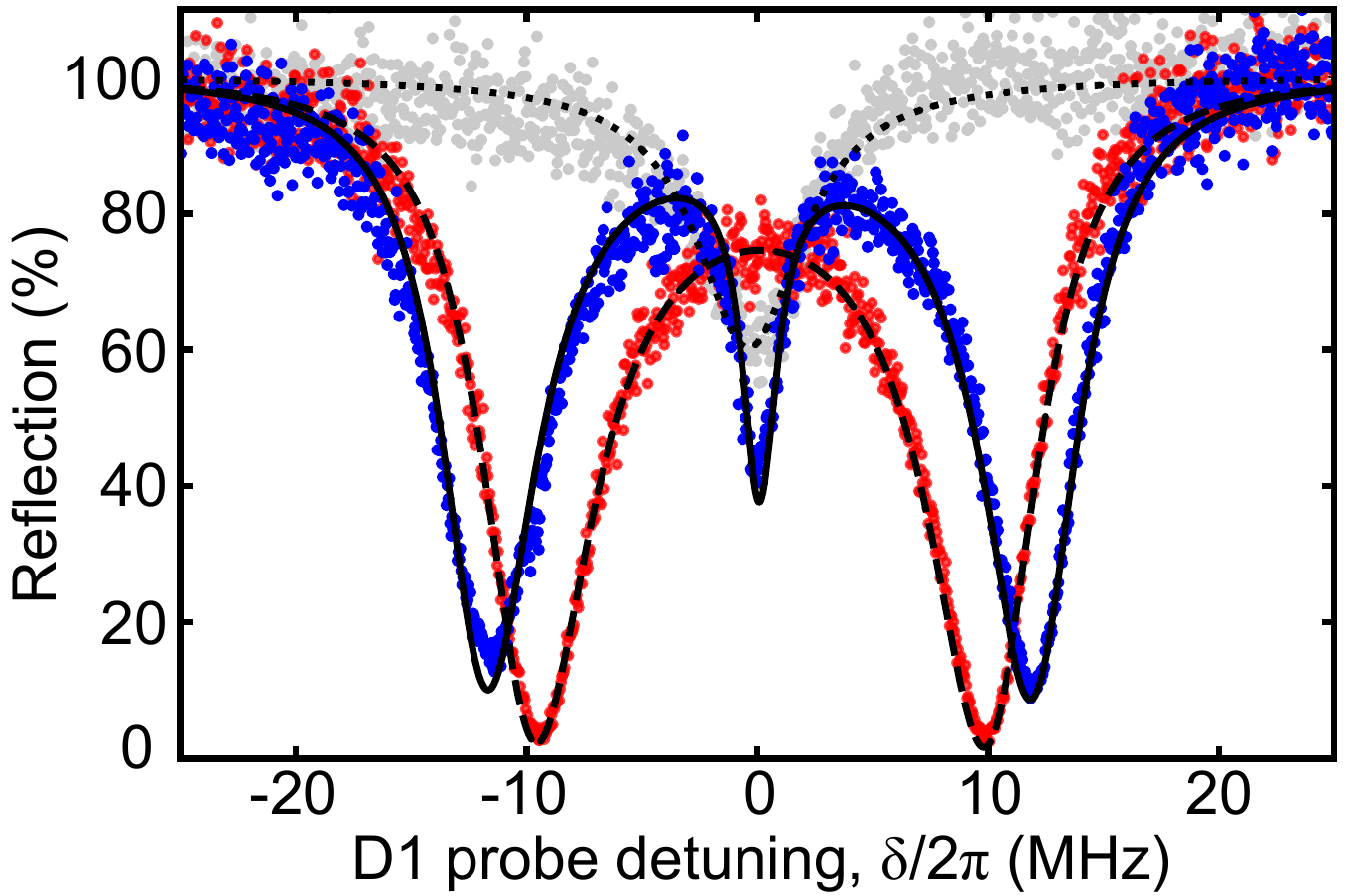}
	\caption{ Reflection spectra: the reflection of the D1 probe, as a function of the probe detuning is measured with an SPCM and normalized to the maximum reflection of the empty cavity (light grey). With the strongly coupled atomic ensemble, we observe the vacuum Rabi splitting of the two reflection dips (red). Adding the EIT control beam, with Rabi frequency $\Omega_c /(2\pi) = \unit{13}{\mega\hertz}$, creates a reflection dip at resonance (blue).}
	\label{fig:ReflSpec}
\end{figure}

Many of the physical parameters of our system are estimated from the EIT transmission and reflection spectra in the weak-excitation regime, where interactions can be neglected. The system can then be modeled by $N$ atoms with a ground state \ket{g=5S_{1/2},F=1,m_F=1}, an intermediate state \ket{e=5P_{1/2},F=2,m_F=2} and a Rydberg state \ket{r=78S_{1/2},J=1/2,m_J=1/2,I=3/2,m_I=3/2}, with respective angular transition frequencies $\omega_{ge}$ and $\omega_{er}$. The $n$-th atom, described by operators $\hat{\sigma}_{ij}^{(n)} = \ket{i}_n\bra{j}_n$ in this basis, is coupled with a constant $g_n$ to the cavity mode. This mode, described by the photon annihilation operator $\hat{a}$, has an angular frequency $\omega_a$ near-resonant with $\omega_{ge}$ and is driven by a weak coherent probe with an amplitude $\alpha$ and an angular frequency $\omega$. A laser beam with an angular frequency $\omega'$ close to $\omega_{er}$ and a Rabi frequency $\Omega_c/(2\pi)$ assumed to be the same for all atoms couples \ket{e} and \ket{r}. In the rotating frame, the Hamiltonian of the system is
\begin{eqnarray}
\nonumber\frac{\hat{H}}{\hbar} &=& -\delta_a\oad\oa -\delta_e\sum_{n=1}^N\os_{ee}^{(n)} -\delta_r\sum_{n=1}^N\os_{rr}^{(n)} \\
\nonumber && +\sum_{n=1}^Ng_n(\oa\os_{eg}^{(n)}+\oad\os_{ge}^{(n)})+\frac{\Omega_c}{2}\sum_{n=1}^N(\os_{er}^{(n)}+\os_{re}^{(n)})\\
 &&+i \alpha\sqrt{2\kappa_0}(\oad-\oa),
\end{eqnarray}
where $\delta_a=\omega-\omega_a$, $\delta_e=\omega-\omega_{ge}$, $\delta_r=\omega+\omega'-\omega_{ge}-\omega_{er}$, and $\kappa_0$ is the field decay rate through the output coupler. Decay terms lead to the master equation
\begin{eqnarray}
\nonumber
\frac{\dd \orho}{\dd t} &=& -\frac{i}{\hbar}[\hat{H},\orho]+\mathcal{D}[\sqrt{2\kappa}\oa]\orho+\sum_{n=1}^N\mathcal{D}[\sqrt{2\gamma}\os_{ge}^{(n)}]\orho\\
&&+\sum_{n=1}^N\mathcal{D}[\sqrt{2\gamma_r}\os_{gr}^{(n)}]\orho,\\
\mathcal{D}[\hat{A}]\orho &=& \hat{A}\orho\hat{A}^\dag-\frac{1}{2}\hat{A}^\dag\hat{A}\orho - \frac{1}{2}\orho\hat{A}^\dag\hat{A}.
\end{eqnarray}
where $\kappa$ is the total cavity field decay rate while $\gamma$ and $\gamma_r$ are the respective decay rates of the coherences $\os_{ge}$ and  $\os_{gr}$. In the  linear response limit $\alpha\rightarrow 0$ we can neglect the populations of the cavity mode and of the excited states. By deriving and solving the steady-state Bloch equations, we obtain the transmission $\mathcal{T}$ (normalized to the bare cavity maximum) and the reflectivity $\mathcal{R}$, involving the complex detunings $\Delta_a = \delta_a + i\kappa$, $\Delta_e = \delta_e + i\gamma$, $\Delta_r = \delta_r + i\gamma_r$  and the collective coupling $g=\sqrt{\sum_n g_n^2}$:
\begin{eqnarray}
\mathcal{T} &=& \left|\frac{\kappa}{\Delta_a-g^2/\left[\Delta_e-\Omega_c^2/(4\Delta_r)\right]}\right|^{2} \\
\mathcal{R} &=& \left|1-\frac{i 2\kappa_0}{\Delta_a-g^2/\left[\Delta_e-\Omega_c^2/(4\Delta_r)\right]}\right|^{2},
\end{eqnarray}
Setting $\Omega_c = 0$ or $g=0$ yields the vacuum Rabi splitted or the Lorentzian bare cavity spectra, respectively.    

These expressions are fitted to the spectra observed with weak probes and presented in Fig.~\ref{fig:setup}(c) for the transmission and in Fig.~\ref{fig:ReflSpec} for the reflection. They provide the values of cavity decay, coupling strength, Rabi frequencies, photon extraction efficiency, transmission and reflectivity given in the main text. Additionally, they provide the intrinsic linewidth of the Rydberg
state $\gamma_r/(2\pi) = \unit{0.12}{\mega\hertz}$. Reducing this linewidth by cooling the atomic motion and compensating stray electric fields with a set of electrodes was essential for achieving a high transparency critical for our applications.


%

\end{document}